\documentclass[11pt]{article}
\usepackage{moriond,epsfig}
\usepackage{amssymb}

\def\Journal#1#2#3#4{{#1} {\bf #2}, #3 (#4)}

\def\be{\begin{equation}}
\def\ee{\end{equation}}
\def\bea{\begin{eqnarray}}
\def\eea{\end{eqnarray}}

\begin{document}
\vspace*{4cm}
\title{SUPERLONG GAMMA-RAY BURSTS
}

\author{Ya. Tikhomirova${}^{1,2}$, B. E. Stern${}^{3,1,2}$}

\address{${}^{1}$ Astro Space Center of Lebedev Physical Institute,
Moscow, Russia;\\
${}^{2}$ Stockholm Observatory, Sweden;\\
${}^{3}$ Institute for Nuclear Research, Moscow, Russia}

\maketitle\abstracts{
Before the BATSE/GRO launch GRBs seem to be a uniform phenomenon 
with duration up to $\sim$100 seconds. The BATSE has detected many events 
longer than 100 s and a few longer than 500s. 
We performed the off-line scan of the 1024 ms continuous BATSE records 
and revealed several non-triggered episodes of the BATSE GRB triggers 
which confidently belong to the same GRBs. There are also several 
pairs of bursts which are candidates to single very long (''superlong'')
GRBs lasting up to $\sim$2000s. Their emission on 500-2000s is prompt
emission rather than afterglow. These superlong GRBs probably
belong to the class of ''long'' GRBs constituting the tail of their
duration distribution. The existence of such events can constrain
some models and should be taken into account in the studies of 
hard X-ray afterglows. 
}

\section{Introduction}
It is presently well known that events starting as Gamma-Ray Bursts
(GRBs) are followed by X-ray and optical afterglows 
(see, for example, \cite{a}).
But the important issue is how long may last the prompt GRB emission.

The most sensitive  GRB experiment was
the Burst And Transient Sources Experiment (BATSE)\cite{BATSE}.
Before BATSE/GRO launch GRBs seem to be a uniform phenomenon
with duration up to $\sim$100 seconds\cite{KONUS}.
The BATSE(1991-2000) detected 4 GRBs with durations 
longer than 500 seconds (see Table~\ref{tab1}).
The longest of them lasted $\sim$1300 seconds.

Recently Valerie Connaughton published\cite{con} the results of her
search for long-lived GRB prompt emission in the BATSE 2048ms
daily records. By summing the signals from hundreds of bursts
she found that tails out to hundreds of seconds after the trigger
may be a common feature of long bursts. Note, that bursts with
precursor or successor emission, defined as 
outbursts which are separated from the time of the peak
by longer than the duration of the longest episode of the burst,
were excluded.

We searched for the superlong prompt emission in the range
500-2000 seconds after the trigger in the BATSE 1024 ms records
for individual GRBs.

\section{Search for superlong GRB prompt emission.}
We performed the off-line scan of the 1024 ms BATSE 
records\cite{stern1,stern2} 
and collected all 3906 found GRBs (both triggered by the BATSE and not) 
to the Uniform Catalog (UC).
We found 10 {\bf ''evident'' superlong events}
(longer than 500 seconds).
9 out of them are the BATSE GRB triggers. 
4 out of 9 are known as very long GRBs
but for 5 out of 9 there are no duration data  
in the BATSE GRB catalog (see Table~\ref{tab1}).
We found also 1 new superlong non-triggered GRB.

\begin{table}
\caption{Statistics of superlong GRBs found in the BATSE data.\label{tab1}}
\vspace{0.4cm}
\begin{center}
\begin{tabular}{|c|c|c|}
\hline
&\multicolumn{2}{|c|}{\bf T90 $>$ 500 s}\\
&500-1000 s&1000-2000 s\\
\hline
The BATSE catalog$^*$(according to the duration table)& 3 & 1 \\
The Uniform Catalog$^{**}$(evident superlong events)& 5 & 1 \\
The search UC for coinciding pairs& 1 & 6 \\
\hline
\multicolumn{3}{c}{$^*$\footnotesize{
available at http://gammaray.msfc.nasa.gov/batse/grb/catalog/current/}}\\
\multicolumn{3}{c}{$^{**}$\footnotesize{
available at http://www.astro.su.se/groups/head/grb\_archive.html}}\\
\end{tabular}
\end{center}
\end{table}

Figure 1 represents the light curves of 2 longest BATSE GRBs.
Both are the BATSE triggers.
We confirm the duration of the longest burst GRB970315 ($\sim$1300 s)
of the complex structure known from the BATSE catalog.
The second longest GRB971208 has the Fast Rise Exponential Decay  (FRED)
light curve and lasts about 1000 s or even more. 
After 1000s there is $\sim$500 s gap in the 1024 ms data we used
and we can only conclude that the duration of this burst is $<$1500 s.
In the BATSE catalog its duration is not estimated.
Both the rise time ($\sim$70 s) and the decay time of GRB971208
are unusually long so this is a unique superlong smooth (but not soft!) 
FRED gamma-ray burst.

\begin{figure}
\psfig{figure=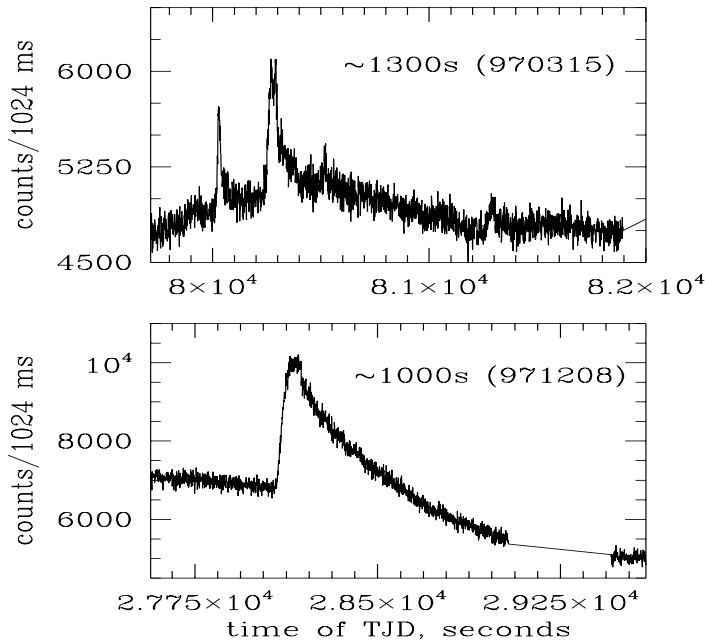,width=5.0in,height=3.5in}
\caption{Time profiles of 2 longest ''evident''GRBs found in
the BATSE 1024 ms records. Count rates are in two brightest detectors
in the  50-300 keV energy range.
In the case of GRB971208  a background subtraction to estimate of 
the duration is problematic because of the gap. 
The duration of GRB970315 t$_{90}$=1307s
according to the BATSE catalog. 
\label{fig:SL1}}
\end{figure}

We checked also all GRBs found in our off-line scan
for pairs which can belong to the same event
({\bf ''coincident pairs''}).
We searched for bursts within $\delta T=2000$ s
with consistent errorboxes.
We found 9 such pairs.
In each case we estimated an  expected number $N$ of
coinciding independent bursts within given location and
time separation for the whole BATSE observation time.
Only for 2 pairs $N>1$ and they were excluded.
Other 7 pair of GRBs were accepted as candidates to superlong GRBs.
As localization of GRB is a difficult task\cite{briggs1999},\cite{stern2}
and typical errorbox is as large as several degrees
some of the pairs could be accidental.
However, the probability of chance observation of all 7 pairs
is roughly a product of their expectations, i.e. $\sim10^{-8}$.

The light curves of 7 coincident pairs are shown in Figure 2.
\begin{figure}[h]
\psfig{figure=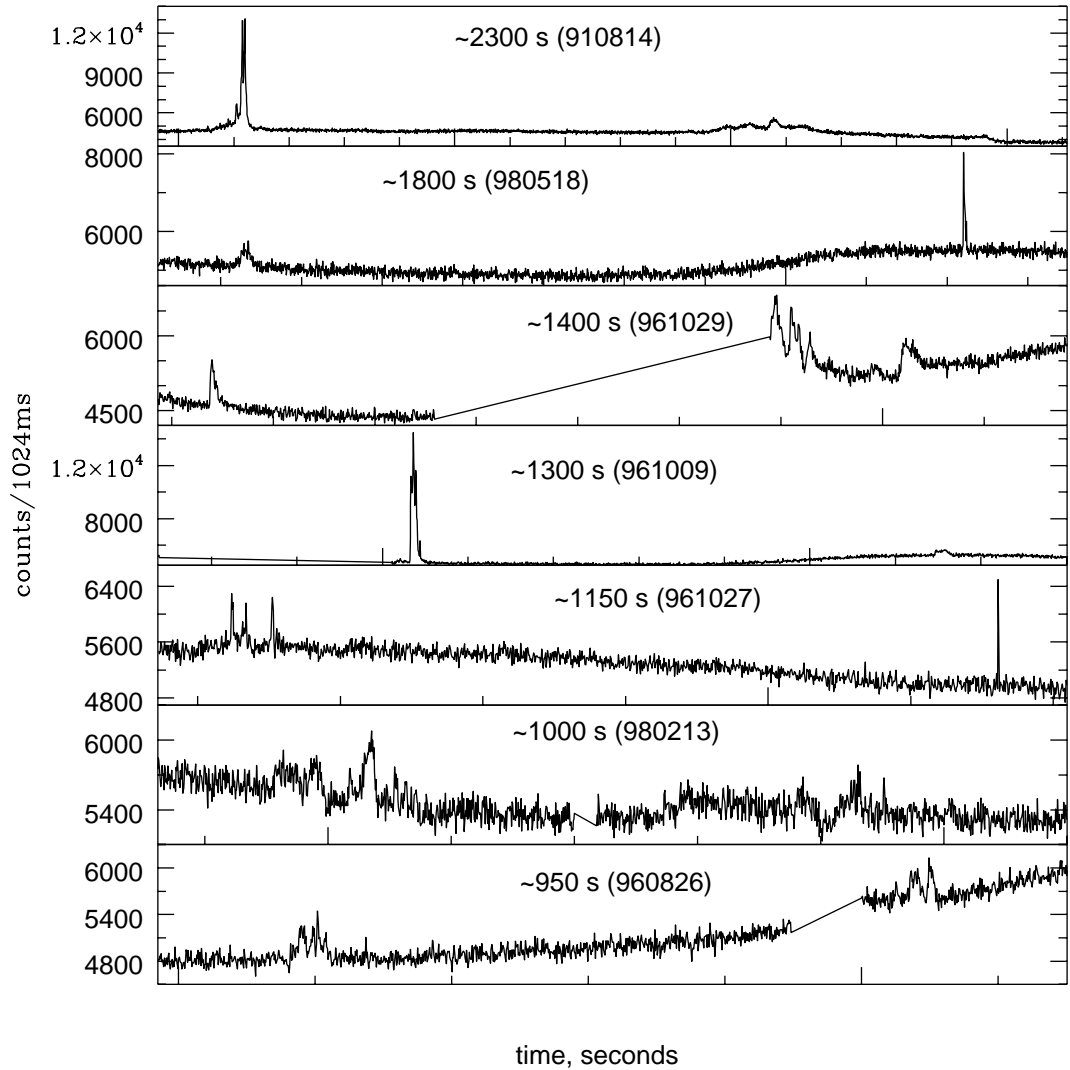,height=6.0in}
\caption{Time profiles of candidates to superlong GRBs
(coincident pairs) found in the BATSE 1024 ms records.
The time interval between X-ticks is 200 s in all panels.
The estimated duration corresponds to the interval
between the start of the first episode and the end of
the last episode.
\label{fig:SL2}}
\end{figure}
The duration between the start of the first episode
and the end of the last episode for the longest event is about 2300 s.
In most cases there is a very long
quiescent interval between episodes.
This is the reason why they were not detected as single GRB.

The time identification, the galactic coordinates,
the radius of the errorbox, the duration, the peak
flux in the 50-300 keV energy range, the hardness as ratio of
the fluence in the  50-300 keV range (2,3 BATSE channels) to that
in the 20-50 keV range (1 ch.) and the expected number $N$ of chance
coincidences of the 7 pairs are listed in Table~\ref{tab:2}.
\begin{table}
\caption{
Superlong ($\gtrsim 1000$ s) GRBs found in the BATSE 1024ms records.\label{tab:2}}
\vspace{0.4cm}
\begin{center}
\begin{tabular}{|c|c|c|c|c|c|c|c|c|c|}
\hline
\hline
Date&Sec.of&Catal./name&$\alpha$,$^o$&$\delta$,$^o$&$R$,$^o$
&$T_{90}$&$F_{peak}$&HR&$N_{cc}$\\
(TJD)&TJD&&&&&$_s$&$_{ph/cm^2/s}$& $_{2,3/1}$&\\
\hline
\hline
\multicolumn{10}{|c|}{\bf 
evident superlong GRB }\\
\multicolumn{10}{|c|}{\bf $\sim$1300 s}\\
{\bf 970315 }&{\bf  80021 }& {\bf UC 10522d }& {\bf 129.9 }
& {\bf -52.6 }& {\bf 3.5 } & {\bf 1307.} & {\bf 0.66} & & {\bf -}\\
(10522) & 80022 & BATSE 6125 & 130.6 & -52.4 & 0.99 & 1307. & no && \\
\hline
\multicolumn{10}{|c|}{\bf $\sim$1000 s }\\
{\bf 971208 }& {\bf 28085 }& {\bf UC 10790a }& {\bf 355.8 }
& {\bf 76.2 }& {\bf 0.4 }& {\bf $>$1000. }& {\bf 2.0 }& {\bf 2.36 }&{\bf -} \\
(10790) & 28092 & BATSE 6526 & 356.5 & 77.9 & 1.2 & - & 1.8 &  & \\
\hline
\multicolumn{10}{|c|}{\bf coincident
pairs as candidates to superlong GRB}\\
\multicolumn{10}{|c|}{\bf $\sim$2300 s}\\
{\bf 910814 }& {\bf 40132 }& {\bf UC 08482c }& {\bf 62.2 }& {\bf 46.8 }
& {\bf 0.8 }& {\bf 72. }& {\bf 4.29 }& {\bf 3.46 }&{\bf 0.06 }     \\
(08482) & 40180 & BATSE 676 & 58.0 & 45.2 & 1.0 & 78. & 4.08 && \\
& {\bf 41941 }& {\bf UC 08482d }& {\bf 60.0 }& {\bf 35.3 }& {\bf 3.1 }
& {\bf 324. }& {\bf 0.58 }& {\bf 2.12 }&\\
&-&-&-&-&-&-&-&&\\
\hline
\multicolumn{10}{|c|}{\bf $\sim$1800 s}\\
{\bf 980518 }& {\bf 22649 }& {\bf UC 10951a }& {\bf 160.3 }& {\bf -44.5 }
& {\bf 11.1 }& {\bf 42. }& {\bf 0.35 }& {\bf 1.69 }& {\bf 0.03 }\\
(10951) &-&-&-&-&-&-&-& &\\
& {\bf 24441 }& {\bf UC 10951b }& {\bf 164.2 }& {\bf -41.9 }& {\bf 5.4 }
& {\bf 8. }& {\bf 1.34 }& {\bf 2.4 }&\\
& 24441 & BATSE 6762 & 162.1 & -42.5 & 2.0 & 8. & 1.46 & &\\
\hline
\multicolumn{10}{|c|}{\bf $\sim$1400 s}\\
{\bf 961029 }& {\bf 23676 }& {\bf UC 10385b }& {\bf 62.1 }& {\bf -53.5 }
& {\bf 6.5 }& {\bf 24. }& {\bf 0.57 }& {\bf 2.34 }& {\bf 0.02 }      \\
(10385) & 23677 & BATSE 5648 & 59.4 & -52.6 & 3.3 & 40. & 0.84 & &\\
& {\bf 24781 }& {\bf UC 10385c }& {\bf 56.4 }& {\bf -53.0 }& {\bf 1.8 }
& {\bf 49. }& {\bf 0.67 }& {\bf 1.68 }&\\
& 24350 & BATSE 5649 & 59.8 & -48.9 & 0.3 & no & no & &\\
\hline
\multicolumn{10}{|c|}{\bf $\sim$1300 s}\\
{\bf 961009 }& {\bf 49065 }& {\bf UC 10365c }& {\bf 135.5 }& {\bf -79.0 }
& {\bf 1.1 }& {\bf 20. }& {\bf 6.69 }& {\bf 2.91 }& {\bf 0.007 }      \\
(10365) & 49065 & BATSE 5629 & 130.2 & -80.2 & 0.4 & no & 6.4 & &\\
&{\bf 50292 }& {\bf UC 10365d }& {\bf 110.2 }& {\bf -79.0 }& {\bf 6.6 }
& {\bf 36. }& {\bf 0.25 }& {\bf 2.57 }&\\
& 50313 & Kom 961009.58 & 80.0 & -80. & 3.9 & 52. & 0.17 & &\\
\hline
\multicolumn{10}{|c|}{\bf $\sim$1150 s}\\
{\bf 961027 }& {\bf 42247 }& {\bf UC 10383d }& {\bf 72.0 }& {\bf -43.5 }
& {\bf 10.4 }& {\bf 75. }& {\bf 0.42 }& {\bf 3.04 }& {\bf 0.2 }      \\
(10383) & 42247 & BATSE 5646 & 67.4 & -42.4 & 5.6 & 109. & 0.47 & &\\
& {\bf 43323 }& {\bf UC 10383e }& {\bf 84.4 }& {\bf -51.0 }& {\bf 17.4 }
& {\bf 2. }& {\bf 0.78 }& {\bf 8.2 }&\\
& 43322 & BATSE 5647 & 68.7 & -54.3 & 5.8 & 1. & 0.85 & &\\
\hline
\multicolumn{10}{|c|}{\bf $\sim$1000 s}\\
{\bf 980213 }& {\bf 63928 }& {\bf UC 10857c }& {\bf 11.1 }& {\bf -23.8 }
& {\bf 7.4 }& {\bf 198. }& {\bf 0.33 }& {\bf 2.23 }& {\bf 0.7 }      \\
{\bf (10857) }& {\bf 64849 }& {\bf UC 10857d }& {\bf 6.2 }& {\bf -10.6 }
& {\bf 13.1 }& {\bf 104. }& {\bf 0.25 }& {\bf 2.7 }&\\
\hline
\multicolumn{10}{|c|}{\bf $\sim$950 s}\\
{\bf 960826 }& {\bf 57175 }& {\bf UC 10321d }& {\bf 191.1 }& {\bf 15.8 }
& {\bf 10.2 }& {\bf 45. }& {\bf 0.32 }& {\bf 1.68 }& {\bf 0.6 }      \\
(10321) & 57204 & Kom 960826.66 & 187.8 & 20.4 & 10.8 & 53. & 0.23 & &\\
& {\bf 58072 }& {\bf UC 10321e }& {\bf 191.1 }& {\bf 28.3 }& {\bf 12.6 }
& {\bf 34. }& {\bf 0.23 }& {\bf 4.63 }&\\
& 58099 & Kom 960826.67 & 179.8 & 21.3 & 5.3 & 40. & 0.38 & &\\
\hline
\end{tabular}
\end{center} 
\end{table}
The same characteristics are given according to
the BATSE catalog (for triggered GRBs) or to the
Untriggered Supplement to the BATSE catalog (for untriggered)\cite{kom}.
The second episodes in these pairs can be softer or harder than
the first, brighter or dimmer.
So, they should be treated as prompt emission rather then afterglow.

It is possible that GRB prompt emission
may last even longer than $\sim$ 2000 s. We could
not identify coincident pairs at $\delta T \gtrsim 2000$ s
because the probability of chance coincidence
for such seperated episodes is too large.
But as we did not detect such evident superlong GRBs
which should be detectable despite of the gaps in the BATSE records
it is evident that GRB prompt emission
longer than $\sim$ 2000 s is rare.

The hardness of these superlong bursts 
(both 2 evident events and 7 coincident pairs)
is typical for GRBs (see Fig. 3 and Tab. 2).
They do not concentrate to the galactic center or
to the galactic plane.
Their light curves  represent the well known rich morphology of
GRB profiles (see Fig. 1,2).
So, we have no arguments to conclude that superlong GRBs are
the separate class of GRBs.

\begin{figure}
\psfig{figure=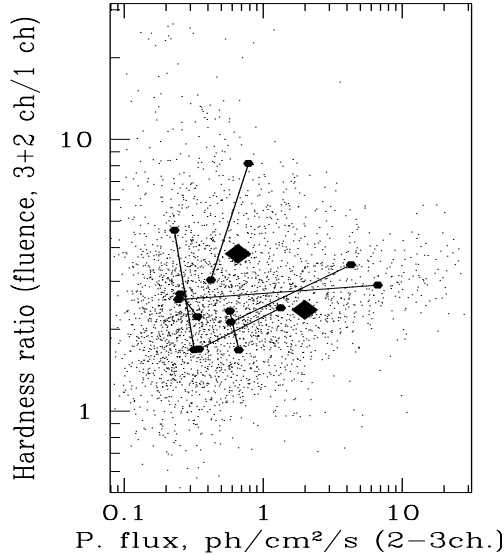,width=4.0in,height=3.0in}
\caption{The hardness of GRBs found in the BATSE 1024 ms records 
as the ratio of fluences 
in the 50-300 keV energy range (2-3 BATSE channels) 
to that in the 20-50 keV (1 ch.) 
Diamonds represent 2 longest evident superlong GRBs.
7 coincident pairs as candidates to superlong GRBs are
shown by connected points. Dots represent all 3906 GRBs
found in the scan of the BATSE records.  \label{fig:SLhr}}
\end{figure}

The duration distribution of GRBs with new data is shown
in Figure 4.
We calculated by Monte-Carlo simulations the probabilities
to detect long and superlong bursts in the BATSE 1024 ms records.
For superlong bursts with long quescent interval it means 
the probability to detect both seperated episodes. 
For evident superlong bursts we estimated the probability
to detect more than 90\% of their duration. 
We took into account regular gaps in the
BATSE data and the coveradge by the Earth of the part of the sky.
The duration distribution was corrected to the estimated probabilities.
For superlong bursts with long quiescent interval
we used the duration between the start of the first episode
and the end of the last episode.
The tail of new distribution is consistent with the power low
slope $\sim$ -1.5.

\begin{figure}
\psfig{figure=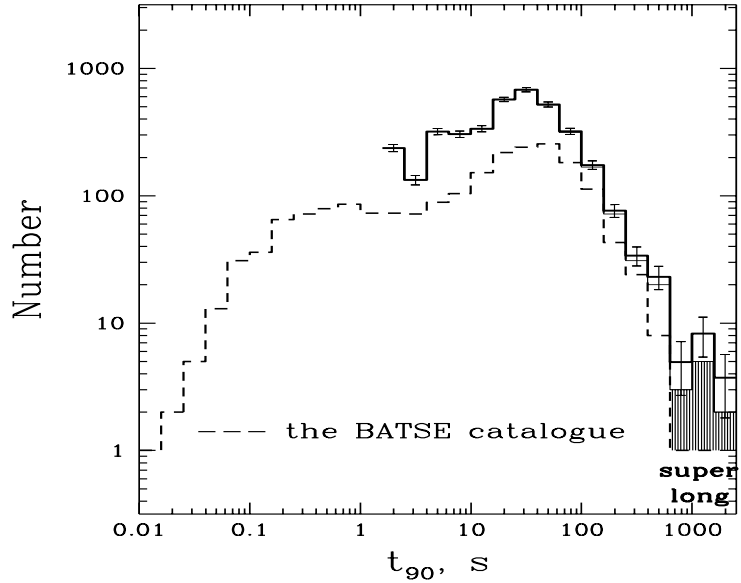,width=5.0in,height=3.5in}
\caption{The duration distribution
of GRBs found in the BATSE 1024 ms records (thin histogram).
The thick histrogram represents the distribution corrected
to the probability to observe the whole duration of a GRB.
The dashed histogram corresponds to the BATSE catalog.\label{fig:DD}}
\end{figure}

\section{Conclusion}
We found evidence that the GRB prompt emission may last in some
cases up to $\sim$2000 seconds.
These are our preliminary results of search for superlong GRB
prompt emission in the BATSE records for individual bursts.
The existence of such superlong GRBs should be explained by
GRB models and taken into account in studies of early
hard X-ray afterglows. 

\section{Acknowledgments}

This work was supported by the Swedish Natural Science Research Council,
the Royal Swedish Academy of Science, and the Wennergren Foundation for
Scientific Research and Russian Foundation for Basic Research
(grants 00-02-16135 and 02-02-06674).

\end{document}